\documentclass[12pt]{article}

\usepackage{amsfonts}
\usepackage{amssymb}
\usepackage{amsbsy}
\usepackage{epsf}

\def\subsubsection{\@startsection{subsubsection}{3}{\z@}{-3.25ex plus
 -1ex minus -.2ex}{1.5ex plus .2ex}{\large\sc}}

\newcommand{\be}{\begin{equation}}
\newcommand{\ee}{\end{equation}}
\newcommand{\bel}[1]{\begin{equation}\label{#1}}
\newcommand{\bea}{\begin{eqnarray}}
\newcommand{\eea}{\end{eqnarray}}
\newcommand{\ba}{\begin{array}}
\newcommand{\ea}{\end{array}}

\newcommand{\bra}[1]{\mbox{$\langle \, {#1}\, |$}}
\newcommand{\ket}[1]{\mbox{$| \, {#1}\, \rangle$}}
\newcommand{\exval}[1]{\mbox{$\langle \, {#1}\, \rangle$}}
\newcommand{\smfrac}[2]{\mbox{\small$\frac{#1}{#2}$}}
\newcommand{\D}{{\rm d}}                

\renewcommand{\vec}[1]{{\boldsymbol{#1}}} 

\newcommand{\zeile}[1]{\vskip #1 \baselineskip} 
\def\bbbz{{\mathchoice {\hbox{$\sf\textstyle Z\kern-0.4em Z$}}
{\hbox{$\sf\textstyle Z\kern-0.4em Z$}}
{\hbox{$\sf\scriptstyle Z\kern-0.3em Z$}}
{\hbox{$\sf\scriptscriptstyle Z\kern-0.2em Z$}}}}
 
\begin{document}
\title{\Large\bf On the universality of the fluctuation-dissipation ratio in
non-equilibrium critical dynamics}
\author{Malte Henkel$^{a}$\footnote{permanent addresses}$^{~,b}$ and 
Gunter M. Sch\"{u}tz$^{a,c*}$ \\[2mm]
\it \small $^{a}$Laboratoire de Physique des 
Mat\'{e}riaux,\footnote{Laboratoire associ\'e au CNRS UMR 7556} Universit\'{e} 
Henri Poincar\'{e} Nancy I,\\ \it \small B.P. 239, 
F--54506 Vand{\oe}uvre-l\`{e}s-Nancy Cedex, France\\[2mm]
\it \small $^{b}$Centro de F\'{\i}sica Te\'orica e Computacional,
Universidade de Lisboa, \\ \it\small Av. Prof. Gama Pinto 2, 
P--1649-003 Lisboa, Portugal\\[2mm]
\it \small $^{c}$ Institut f\"ur 
Festk\"orperforschung, Forschungszentrum 
J\"ulich,\\ \it \small D--52425 J\"ulich, Germany}
\date{\today}
\maketitle

\begin{abstract}
The two-time nonequilibrium correlation and response functions 
in $1D$ kinetic classical spin systems with non-conserved dynamics and 
quenched to their zero-temperature critical
point are studied. The exact solution of the kinetic Ising model with
Glauber dynamics for a wide class of initial states allows for an
explicit test of the universality of the non-equilibrium 
limit fluctuation-dissipation ratio $X_{\infty}$. It is shown that the value of
$X_{\infty}$ depends on whether the initial state has finitely many domain
walls or not and thus two distinct dynamic universality classes can be 
identified in this model. Generic $1D$ kinetic spin systems with 
non-conserved dynamics fall into the same universality classes as the 
kinetic Glauber-Ising model provided the dynamics is invariant under the
C-symmetry of simultaneous spin and magnetic-field reversal. While C-symmetry
is satisfied for magnetic systems, it need not be for lattice gases which
may therefore display hitherto unexplored types of non-universal kinetics. 
\end{abstract}

\newpage

\section{Introduction}

The understanding of the long-time behaviour of strongly interacting
systems with many degrees of freedom and evolving far from equilibrium 
is an active topic of much current interest, 
see \cite{Cate00,Godr02,Cugl02,Cris03}
for recent reviews. Besides the more far-reaching
aspects of disordered systems undergoing glassy behaviour, many of the 
fundamental questions of non-equilibrium statistical mechanics can already
be studied on the conceptually simpler kinetic ferromagnetic systems. In
several instances, general ideas can be subjected to exacting tests because
several non-trivial and exactly soluble models are available. 

In this paper, we consider the kinetics of a purely classical 
spin system with a nonconserved order parameter and an equilibrium
critical temperature $T_c\geq 0$ and quenched to a final temperature 
$T\leq T_c$ from some initial state. Then fluctuations of the
initial state will lead on the microscopic level to the growth of 
correlated domains and the slow movement of the domain boundaries will drive
the irreversible time-evolution of the macroscopic observables. 
It has been realized in
recent years that the associated ageing effects are more fully revealed
through the study of {\em two-time} correlation functions $C(t,t_w)$ 
and response functions $R(t,t_w)$ (see section 2 for the precise definitions)
and where $t$ is referred to as observation time while $t_w$ is called the
waiting time.  In addition, it has been understood that correlation and 
response functions must be studied together, since out of equilibrium the
fluctuation-dissipation theorem is no longer valid. It is usual to
characterize the breaking of the fluctuation-dissipation theorem in terms
of the {\em fluctuation-dissipation ratio} (FDR) \cite{Cugl94}
\be
X(t,t_w) := T R(t,t_w) \left(\frac{\partial C(t,t_w)}{\partial t_w}\right)^{-1}
\ee
where at equilibrium, one would recover $X(t,t_w)=1$ (at $T=0$, some care is 
needed to absorb $T$ into the definition of the response function). 
In the scaling limit, where simultaneously $t_w$ {\em and} $t-t_w$ become
large, one actually finds that the fluctuation-dissipation ratio
$X(t,t_w)=\hat{X}(t/t_w)$ satisfies a simple scaling law in terms of the
scaling variable $x=t/t_w$, see \cite{Cris03} for a recent review. 
A quantity of particular interest is the {\em limit fluctuation-dissipation
ratio} $X_{\infty}$ defined by
\be
X_{\infty} = \lim_{t_w\to\infty} \left( \lim_{t\to\infty} X(t,t_w) \right) = 
\lim_{x\to\infty} \hat{X}(x)
\ee
For a quench into the disordered phase, it is known that $X_{\infty}=0$ in
general. But for a critical quench with $T=T_c$, 
Godr\`eche and Luck \cite{Godr00,Godr00b} have proposed that 
$X_{\infty}$ should be an {\em universal} quantity. 

The evidence supporting this conjecture (which in this paper we shall
call {\em `universality'} for short) was based on the available results coming
from exactly solvable kinetic spin systems quenched from a fully disordered
state and from simulations in the $2D$ and $3D$ kinetic Ising model with
Glauber dynamics, see \cite{Godr02} for a review. Further supporting evidence
in favour of the universality conjecture comes from field-theoretic two-loop
calculations of the O($n$)-model, again starting from a fully disordered
initial state \cite{Cala02a,Cala02b}. The universality of $X_{\infty}$ has
also been confirmed numerically for the $2D$ Glauber-Ising and voter models
\cite{Sast03}. 

Statements about the universality of a physical quantity are best tested by
varying important control parameters of suitable models and then studying their
effects. Indeed, the r\^ole of spatially long-ranged initial correlations
of the form
\bel{1-3} 
C_{\rm ini}(\vec{r}) \sim |\vec{r}|^{-\nu}
\ee
where $\nu$ is a control parameter, was studied in the kinetic 
spherical model at $T=T_c$ \cite{Pico02}. It was shown that there exists an 
unexpectedly rich kinetic phase diagram, depending on $\nu$ and the space 
dimension $d$. In most of these phases, either $X_{\infty}=0$ or else it is 
independent of $\nu$, but in one
phase $X_{\infty}$ was shown to be a function of $\nu$ \cite{Pico02},
thus furnishing an important qualification against the 
unrestricted universality of $X_{\infty}$.\footnote{In this phase both the
space dimension $d$ as well as the effective dimension $D=2+\nu$ of the
initial correlations are below the upper critical dimension $d^*$.} 
At present, it is not clear yet whether these results might not simply reflect
a peculiarity of the spherical model. Therefore, we shall study here the
non-equilibrium critical dynamics of $1D$ ferromagnetic spin systems quenched
to their critical temperature $T=0$. 

In order to get analytical results, we consider in section 2 the exactly
solvable Glauber-Ising model. We shall show that for initial correlations of
the form (\ref{1-3}) with $\nu\geq 0$, we have indeed universality of
the entire function $\hat{X}(x)$, and thus in particular of $X_{\infty}$, in 
agreement with the Godr\`eche-Luck conjecture. However, in section 3 we study 
even more general initial states which consist of large ordered domains and 
then show that the scaling function $\hat{X}(x)$ as well as $X_{\infty}$ are 
different from the ones obtained for initial states of the form
(\ref{1-3}). We thereby identify two dynamic universality classes. 
These results are extended in section 4 to generalized two-state 
spin systems evolving according to a non-conserved dynamics with detailed 
balance. We show that those which satisfy a certain global symmetry (which
we call {C-symmetry}) are in one of the two universality classes found for the
Glauber-Ising model. We also comment on the fact that in lattice gases, 
there is no need to satisfy C-symmetry which may lead to further 
non-universalities. We conclude in section 5.

\section{Glauber dynamics with correlated initial conditions}

\setcounter{equation}{0}

We shall study the Ising chain with equilibrium Hamiltonian
${\cal H}=-J\sum_n s_n s_{n+1}$ where $s_n=\pm 1$ denotes the Ising spins and
$J>0$ is the interaction strength. We consider translation-invariant initial 
distributions such that the initial magnetization $m_0 = \exval{s_n}$ should
be different from $\pm 1$, i.e. there is a {\em finite density} of domain walls
at initial time. In the literature usually symmetric initial distributions 
with $m_0=0$ are considered. In this section, we allow for general non-symmetric 
initial distributions with $-1 < m_0 <1$. The case $|m_0|=1$ requires separate 
treatment and is studied in section 3. In the absence of a magnetic
field we assume Glauber dynamics \cite{Glau63} for the stochastic time 
evolution of the spins. We set the time scale for individual spin flips to
unity.

\subsection{Two-time correlation function}

It is convenient to write the two-time correlation function
\be
C_n(t+\tau,t) = C_n(t;\tau) := \exval{s_n(t+\tau)s_0(t)}
\ee
in the quantum Hamiltonian formalism, see \cite{Schu00,Henk03} for recent 
reviews. One has
\bel{2-1}
C_n(t;\tau) = \bra{s} \sigma^z_n \mbox{e}^{-H_0\tau} \sigma^z_0 
\mbox{e}^{-H_0 t}\ket{P_0}
\ee
where $\sigma^{x,y,z}_n$ are the Pauli matrices acting on the $n^{\rm th}$ site 
of the chain and
\bel{2-7}
H_0 = \frac{1}{2} \sum_n (1-\sigma^x_n)\left( 1-\frac{\gamma}{2}\sigma^z_n
(\sigma^z_{n-1}+ \sigma^z_{n+1})\right)
\ee 
is the Markov generator (stochastic Hamiltonian) for Glauber dynamics 
\cite{Feld71}. Furthermore, $\ket{P_0}$ is the probability vector representing 
the initial distribution of spins and the constant summation vector $\bra{s}$ 
is the left steady state. The generator is constructed such that it
satisfies detailed balance with respect to the equilibrium distribution at 
temperature $T$ of the $1D$ zero-field ferromagnetic Ising model with 
interaction strength $J$. This is achieved by setting 
$\gamma = \tanh{(2J/T)}$ (we use units such that the Boltzmann constant 
$k_{\rm B}=1$). We introduce the shorthand 
\be 
C_n := C_n(0,0) \;\; ; \;\; C_n(t) := C_n(t;0) = \exval{s_n(t)s_0(t)}
\ee
for the initial correlations and for the equal-time 
correlation function, respectively. Notice that  
\bel{2-2}
C_n(t) = C_{-n}(t)\;\; ; \;\; C_0(t)=1 \quad \forall t\geq 0.
\ee

For Glauber dynamics the time-evolution of the spin-expectation is linear 
and at vanishing temperature $T=0$ satisfies a lattice diffusion equation for 
a $1D$ random walk with hopping rate 1/2 \cite{Glau63}. The propagator 
\bel{2-3}
G_n(y) = \mbox{e}^{-y} I_n(y)
\ee 
of the lattice diffusion equation involves the
modified Bessel function $I_n(y)$ \cite{Grad81} and describes the probability 
of moving a distance of $n$ lattice units after time $y$. Hence
\bel{2-4}
\bra{s} \sigma^z_n \mbox{e}^{-H_0\tau} = \sum_{m=-\infty}^{\infty} 
\mbox{e}^{-\tau} I_{n-m} (\tau) \bra{s} \sigma^z_m
\ee
which immediately yields
\bel{2-5}
C_n(t;\tau) = \sum_{m=-\infty}^{\infty} \mbox{e}^{-\tau} 
I_{n-m} (\tau) C_m(t).
\ee

It remains to calculate the equal-time two-point correlation function $C_m(t)$.
{}For special initial distributions this has already been done in the classical 
paper by Glauber \cite{Glau63}. For our more general treatment we observe
that the total correlation function may be split into an interaction 
part $C_m^{\rm int}(t)$ and a correlation part $C_m^{\rm corr}(t)$. The latter 
one vanishes for uncorrelated (infinite-temperature) initial states. Then
\bea
C_m(t) & = & \mbox{e}^{-2t} I_m(2t) + 2 \sum_{k=1}^{\infty} 
\mbox{e}^{-2t} I_{|m|+k}(2t) \nonumber \\
\label{2-6}
 & & + \sum_{k=1}^{\infty} \mbox{e}^{-2t} 
       \left[I_{|m|-k}(2t)-I_{|m|+k}(2t)\right] C_k \\
 & =: & C_m^{\rm int}(t) + C_m^{\rm corr}(t). \nonumber
\eea
The two-time autocorrelation function then follows from (\ref{2-5}) by setting
$n=0$.

The Laplace transform of the interaction part of the two-time autocorrelation 
function has already been studied in detail by Godr\`{e}che and 
Luck \cite{Godr00} and by Lippiello and Zannetti 
\cite{Lipp00}.\footnote{Multispin correlators and associated response functions
of the $1D$ Glauber-Ising model have been studied recently 
in \cite{Maye03a,Maye03b}. Two-time correlators of the Ising chain in a 
transverse field are calculated in \cite{Subr03}.} Here, we 
prefer to work in the more intuitive time-domain, which allows for an easier
analysis of more general initial conditions. From (\ref{2-6}), we also
define $C_n^{\rm ini}(t;\tau)$ and $C_n^{\rm corr}(t;\tau)$ which will
be calculated separately and $C_n(t;\tau)=C_n^{\rm ini}(t;\tau)+
C_n^{\rm corr}(t;\tau)$. Using the completeness 
property $\sum_n G_n =1$ of the lattice propagator we split the interaction 
part into a contribution which is large at early times and a second 
contribution which dominates the late-time behaviour. We find
\bea
\label{2-8}
C^{\rm int}_0(t;\tau) & = & \mbox{e}^{-(\tau+2t)} I_0(\tau+2t) + 
\mbox{e}^{-\tau}I_0(\tau)
\left[1- \mbox{e}^{-2t}I_{0}(2t)\right] \nonumber \\
 & & + 4 \sum_{m=1}^{\infty}\sum_{k=1}^{\infty} \mbox{e}^{-2t} 
       I_{|m|+k}(2t)\mbox{e}^{-\tau}I_m(\tau).
\eea
For $t,\tau \gg 1$ only the late-time part (containing the double sum) 
plays a role. Using the asymptotic Gaussian form
\bel{2-9}
G_n(y) = \frac{1}{\sqrt{2\pi y}} \mbox{e}^{-\frac{n^2}{2y}}
\left( 1 + {\rm O}\left( y^{-1}\right)\right)
\ee
of the lattice propagator we can turn the sums into integrals. Setting
\bel{2-10}
\alpha = \sqrt{\frac{\tau}{2t}}
\ee 
we obtain
\bea
C^{\rm int}_0(t;\tau) & = & \frac{4}{\pi} \int_0^{\infty} \!\D u 
\int_0^{\infty} \!\D v\,
 \mbox{e}^{-u^2 + (u\alpha + v)^2} \nonumber \\
\label{2-11}
 & = & 1 - \frac{2}{\pi} \arctan{\alpha} = \frac{2}{\pi}\, 
\arctan{\frac{1}{\alpha}}.
\eea
The correlation function depends only on the scaling variable $\alpha$.

For the later comparison with the response function, we now make a small
change in notation, in order to be compatible with the notations usually
employed \cite{Godr00,Pico02,Lipp00}. The quantity denoted `time' $t$ so far,
we shall from now on call {\em waiting time} $s=t_w$. In turn, we call
$t_w+\tau$ {\em observation time} and write $t:=t_w+\tau$. We repeat the
correlation function in this notation, in agreement with \cite{Godr00,Lipp00}
\bel{2-11bis}
C_0^{\rm int}(t,t_w) = \frac{2}{\pi}\arctan\sqrt{\frac{2t_w}{t-t_w}}
\ee
For later use, we also record the derivative of $C_0^{\rm int}$ with
respect to $t_w$. Keeping $t$ fixed, we have 
\bel{2-dcorr}
- \frac{\partial}{\partial t_w} C^{\rm int}_0(t,t_w)
= -\frac{1}{2\pi}\frac{1+2\alpha^2}{\alpha(1+\alpha^2)}\cdot\frac{1}{t_w}
= \frac{\sqrt{2}}{\pi}\frac{x}{\sqrt{x-1\,}(x+1)}\cdot\frac{1}{t_w} 
\ee
for either the scaling variable $\alpha$ or $x=t/t_w$. 

We still have to analyze the correlation part of the autocorrelation function
which has not been studied previously. In order to do so, we use the integral 
representation
\bel{2-13}
G_n(y) =   \frac{1}{2\pi} \int_{-\pi}^\pi \!\D q\, \cos{(qn)} 
\mbox{e}^{-\epsilon_q y}
\ee
of the lattice propagator with the dispersion relation 
$\epsilon_q=(1-\cos{q})$. This yields the exact expression
\bea
\label{2-14}
C^{\rm corr}_0(t_w;\tau) & = & \sum_{m=-\infty}^{\infty} \mbox{e}^{-\tau} 
I_m(\tau) C^{\rm corr}_m(t_w)  
\nonumber \\
 & = & \sum_{m=1}^{\infty} G_m(\tau)
\sum_{n=1}^{\infty} C_n
\frac{2}{\pi} \int_{-\pi}^\pi \!\D q\, \sin{(qm)}\sin{(qn)} 
\mbox{e}^{-\epsilon_q t_w}. ~~~
\eea
For $t_w,\tau \gg 1$ the sum over $m$ may be turned into an integral and the 
asymptotic expression (\ref{2-9}) can be used. After some algebra we find
\bel{2-15}
C^{\rm corr}_0(t_w;\tau)= \frac{4\alpha}{\pi^{3/2}t_w^{1/2}} 
\sum_{n=1}^{\infty} C_n 
 \int_0^{\pi} \!\D q\, q \sin{\left(\frac{q n}{\sqrt{t_w}} \right)}
\, {}_1F_1({\small \smfrac{1}{2}};{\small \smfrac{3}{2}};\alpha^2 q^2) 
\mbox{e}^{-q^2(1+\alpha)^2}
\ee
where ${}_1F_1(a;b;x)$ is a confluent hypergeometric series \cite{Grad81}. 
In order to analyze the leading behaviour for waiting times $t_w\gg 1$ we have 
to distinguish three cases, which depend on the form of the initial 
correlation. We shall in this section consider the following form
\bel{2-18}
C_n(0,0) = C_n \sim \frac{B}{n^\nu} \quad \mbox{for } n \to \infty
\ee
where $\nu\geq 0$ and $B$ are -- {\it a priori} nonuniversal -- control 
parameters. We also define the unnormalized first moment
\bel{2-16}
A := \sum_{n=1}^\infty n C_n
\ee
of the initial correlation function. We can now identify three distinct
situations. 
\\[4mm]
\noindent \underline{Case 1: $A<\infty$}\\[4mm]
Consider first the situation when the series $A$ converges
to a {\em finite} value. Physical examples might be antiferromagnetic 
alternating-sign correlations or for rapidly decaying ferromagnetic 
correlations with $\nu>2$. For large waiting times
$t_w$ the leading contribution to the integral in (\ref{2-15}) comes from small
values of the integration variable $q$ (which describe the long wave-length
fluctuations of the spin system). We may then expand the
sine in its Taylor series and the integration can be performed explicitly
from the series representation of the hypergeometric function. Summing up the
infinite series of Gaussian integrals yields the surprisingly simple exact
asymptotic expression, to leading order in $t_w$
\bel{2-17}
C^{\rm corr}_0(t_w,\tau)= \frac{A}{\pi} \frac{\alpha}{1+\alpha^2}
\cdot \frac{1}{t_w}.
\ee
Because of the extra factor $1/t_w$, this is asymptotically smaller 
than the interaction part (\ref{2-11}).
Hence the nonuniversal amplitude $A$ does not contribute to the leading 
late-time asymptotics of the two-time autocorrelation function.\\[4mm]
\noindent \underline{Case 2: Slowly decaying ferromagnetic 
correlations}\\[4mm]
We now consider the initial correlator (\ref{2-18}) with $0 < \nu < 2$ 
(such that $A$ does not converge). Replacing the summation 
over $n$ by an integration
\typeout{*** Hier ist ein Seitenvorschub ! ***}
\newpage
\bea
\label{2-19}
\frac{1}{\sqrt{t_w}} \sum_{n=1}^\infty C_n \sin{\left(\frac{q n}{\sqrt{t_w}} 
\right)}
& \to & \int_0^\infty \!\D y\, C(y\sqrt{t_w}) \sin{qy} \\
\label{2-20}
& = & B t_w^{-\nu/2} \Gamma{(1-\nu)} \cos{\left(\frac{\pi\nu}{2}\right)} 
|q|^{\nu-1} \mbox{sign}{(q)}. ~~~~~
\eea
This yields
\bel{2-21}
C^{\rm corr}_0(t_w;\tau)= \frac{B 2^{1-\nu}}{\pi}\,
\Gamma\left(1-\frac{\nu}{2}\right)\, 
\frac{\alpha}{(1+\alpha^2)^{(1+\nu)/2}} 
{}_2F_1({\smfrac{1}{2}},{\smfrac{1+\nu}{2}};{\smfrac{3}{2}};
{\small\frac{\alpha}{1+\alpha^2}}) \cdot t_w^{-\nu/2}.
\ee
Also in this case the correlation part of the two-time autocorrelation 
function is asymptotically small compared to the interaction part. 
We conclude that the non-universal 
quantities $B$ and $\nu$, which characterize the initial distribution before 
the quench, do not enter into the leading late-time behaviour.\\[4mm]
\noindent \underline{Case 3: Partial ferromagnetic long range order}\\[4mm]
{}The case $\nu=0$ corresponds to an asymptotically constant spin-spin 
correlation function in the initial state, mimicking (partial) ferromagnetic 
long range order. Such initial states may for example be obtained by quenching
from a uniformly magnetized initial state to zero temperature and zero field 
and had already been studied in \cite{Pico02} where $B=m_0^2$ was related to 
the initial magnetization.
Performing the same steps as in case 2 we find
\bel{2-22}
C^{\rm corr}_0(t_w;\tau)=  \frac{2B}{\pi}\, \arctan{\alpha}.
\ee
This is of the same order of magnitude as the interaction
contribution to the correlation function. For the total correlation function
we obtain
\bel{2-23}
C_0(t_w;\tau) = 1 - (1-B)  \frac{2}{\pi}\, \arctan{\alpha}.
\ee
and therefore, for $t$ fixed we have from (\ref{2-dcorr})
\bel{2-24}
- \frac{\partial}{\partial t_w} C_0(t,t_w) = 
\frac{1-B}{\pi t_w}\frac{x}{1+x}
\sqrt{\frac{2}{x-1}}
\ee
in terms of the scaling variable $x = t/t_w$. This is of the same form as
in the uncorrelated case, but with a nonuniversal amplitude $1-B$ 
determined by the initial long range order. The form (\ref{2-24}) proves
universality with regard to details of the initial distribution
of the two-time correlation function, except for a nonuniversal
amplitude.

\subsection{Two-time response function}

Now we consider the time evolution of the local magnetization 
\bel{2-25}
S_n(t) = \exval{\sigma^z_n(t)} = \bra{s} \sigma^z_n \mbox{e}^{-Ht} \ket{P_0}
\ee
for an initial distribution with initial magnetization $m_0\neq\pm 1$. In zero 
field $S_n(t)=m_0$ for $T=0$ Glauber dynamics. 
To study the linear response of the system to a small localized
perturbation by a magnetic
field we let an external field act at site 0 of the lattice. In the quantum 
Hamiltonian formulation
this perturbation of the zero-field dynamics is represented by the perturbed
Markov generator
\bel{2-26}
H = H_0 + V(h).
\ee
The perturbation $V(h)$ is determined by the requirement that the full 
generator $H$ satisfies detailed balance with respect to the equilibrium
distribution 
\bel{2-27}
P^\ast \sim \exp{\left[\frac{1}{T}
\sum_n \left(Js_n s_{n+1} + h s_0\right)\right]}
\ee
of the ferromagnetic Ising system with interaction strength $J$ and local 
magnetic field $h$ at site 0. This requirement, on which the usual equilibrium 
fluctuation-dissipation theorem is based, does not uniquely fix $V$, as 
different dynamical rules may lead to the same equilibrium distribution 
(\ref{2-27}). In \cite{Godr00} a heat bath prescription was used to implicitly
define $V$. Here we follow more closely the philosophy of Glauber
\cite{Glau63} and define a minimally perturbed dynamics by
\bel{2-28}
V = \frac{1}{2}\left(1-\sigma_0^x\right)
\left[1-\frac{\gamma}{2}
\sigma^z_{0}(\sigma^z_{-1}+\sigma^z_{1})\right]\left[\mbox{e}^{-(h/T)
\sigma^z_{0}}-1\right].
\ee
At zero temperature one has $\gamma=1$. We
shall use the dimensionless field strength $h/T$ throughout this work.

{}Following standard procedures we let the field act at time $t_w$ and
calculate the linear response function (in units of $T$)
\bel{2-29}
R_n(t,t_w) = R_n(t_w;\tau) = \frac{\delta}{\delta h(t_w)} S_n(t)
\ee
at observation time $t$. As before $\tau = t-t_w \geq 0$ is the time elapsed
after the perturbation. By expanding the full time evolution operator
$\exp{(-Ht)}$ in powers of $h$ we find from (\ref{2-25}), (\ref{2-29})
\bel{2-30}
R_n(t_w;\tau) = - \bra{s} \sigma^z_n \mbox{e}^{-H\tau} V' 
\mbox{e}^{-Ht_w}\ket{P_0}.
\ee
Here $V'$ is the derivative of $V$ with respect to $h/T$ taken at $h=0$.
Using (\ref{2-4}) we see after a little algebra that the autoresponse 
function ($n=0$) factorizes
\bel{2-31}
R_0(t_w;\tau) =  \mbox{e}^{-\tau}I_0(\tau) \left[1- C_1(t_w)1\right]
\ee
into the autopropagator $G_0(\tau)$ and a contribution involving the two-point
correlation function at time $t_w$.

To calculate the interaction part of the response function we deduce by
analogy with (\ref{2-8}) 
\bel{2-32}
1-C^{\rm int}_1(t) =  \mbox{e}^{-2t}(I_0(2t)+I_1(2t)).
\ee
For large times $t_w\gg 1$ we then find
\bel{2-33}
R_0^{\rm int}(t_w;\tau) = \frac{1}{\sqrt{2\pi \tau}}\frac{2}{\sqrt{\pi t_w}}
=\frac{1}{\pi t_w}\frac{1}{\alpha}.
\ee
With $\alpha^2=(x-1)/2$ one obtains the autoresponse function in terms of the
scaling variable $x=t/t_w$. Interestingly, the same asymptotic result was found 
in \cite{Godr00} for heat bath dynamics.

The calculation of the correlation part of the autoresponse function proceeds
along the same lines as the calculation of the autocorrelation function. We 
find\\[4mm]
\noindent \underline{Case 1: $A<\infty$}\\[4mm]
Here
\bel{2-34}
C^{\rm corr}_1(t_w) = \frac{A}{4\pi^{1/2}t_w^{3/2}}
\ee
for large $t_w$. Comparison with (\ref{2-32}) shows that we have a subleading 
behaviour of the correlation contribution.\\[4mm]
\noindent \underline{Case 2: Weakly decaying ferromagnetic 
correlations}\\[4mm]
Computing the correlation as above leads to
\bel{2-35}
C^{\rm corr}_1(t_w) = \frac{B}{2^{\nu}\sqrt{\pi}}
\Gamma\left(1-\frac{\nu}{2}\right)\, 
t_w^{-(\nu+1)/2},
\ee
which corresponds again to subleading behaviour. Hence initial correlations 
decaying to zero do not change the asymptotic behaviour of the autoresponse 
function.\\[4mm]
\noindent \underline{Case 3: Ferromagnetic long range order}\\[4mm]
{}For $\nu=0$ (correlations decaying to a constant value $B$) we obtain
\bel{2-36}
C^{\rm corr}_1(t_w) = \frac{B}{\sqrt{\pi t_w}}
\ee
which is of the same order as the interaction part. Therefore
\bel{2-37}
R_0(t_w;\tau) = \frac{(1-B)}{\pi\sqrt{2 \tau t_w}} = \frac{1-B}{2\pi t_w}
\frac{1}{\alpha}.
\ee
We see from equations (\ref{2-24}), (\ref{2-37}) that the same nonuniversal 
amplitude enters the two-time correlation function and the response function 
respectively. 

We can now state the main result of this section: 
in each of the cases 1-3 the fluctuation-dissipation ratio 
$X=R/\dot{C}$ does not depend on the initial state and is given by
\bel{2-38}
X(t,t_w) = R(t,t_w)\left(\frac{\partial C(t,t_w)}{\partial t_w}\right)^{-1}
= \hat{X}(x) = \frac{x+1}{2x}.
\ee
The same result was obtained in \cite{Godr00} for different microscopic 
dynamics and uncorrelated initial states.
We note in passing that one may easily check that also for complete initial 
antiferromagnetic order $C_n = (-1)^n$ the FDR has the same asymptotic form. 
In the limit $x\to\infty$ we find
\bel{2-39}
X_{\infty} = \lim_{x\to\infty} \hat{X}(x) = \frac{1}{2}
\ee
in full agreement with the conjecture \cite{Godr00,Godr00b} that $X_{\infty}$
is an universal constant. 

\section{Low-temperature initial states}
\setcounter{equation}{0}

In the previous section we assumed a translation invariant state with a 
{\it finite density} of domain walls in the
Ising system at the initial time. However, at very low temperatures it is
more relevant to study the time evolution of an almost ordered system with
only {\it finitely many} domain walls at the initial time. For definiteness
we consider two domain walls located at sites $-L$ and $L$ respectively
of the lattice. This corresponds to the initial configuration 
\bel{3-1}
P_0 = \dots \downarrow
\downarrow\downarrow \uparrow \uparrow \dots  \uparrow\uparrow
\downarrow\downarrow\downarrow\dots
\ee
where the inversions of the spins occurs at the positions $-L$ and $+L$,
respectively. Although these initial conditions break translation invariance,
we have chosen the coordinate system such that reflection symmetry with 
respect to the origin is maintained. This is not crucial, but simplifies 
some of the exact expressions to be derived below.

In order to calculate the two-time correlation function we use the 
enantiodromy relation 
\bel{3-2}
H_0^T = B H_A B^{-1}
\ee
between Glauber dynamics and diffusion-limited pair annihilation (DLPA)
\cite{Henk95,Schu00}.\footnote{This relation is not to be confused with the 
duality relation \cite{Sant97a} between Glauber dynamics and diffusion-limited 
pair annihilation.} The process DLPA describes independent random walkers 
hopping with rate $1/2$ to a nearest-neighbour site and annihilating 
instantaneously upon encounter. Here $H_0^T$ is the transpose of the 
Hamiltonian for zero-temperature Glauber
dynamics, $H_A$ is the Markov generator for DLPA 
and $B$ is the factorized similarity transformation
$B=b^{\otimes N}$ with the local transformation matrix
\bel{3-3}
b = \left( \ba{rr} 1 & -1 \\
                   0 &  2 \ea \right).
\ee  
With the enantiodromy relation (\ref{3-2}) and  the initial state (\ref{3-1}) 
one obtains in the thermodynamic limit $N\to\infty$
\bel{3-4}
C_{m,n}(t) = \bra{s} \sigma^z_m \sigma^z_n \mbox{e}^{-H_0t} \ket{P_0} = 
\bra{s} \prod_{k=-L}^L \sigma^z_k\: \mbox{e}^{-H_A t} \ket{m,n}.
\ee
By identifying spin up with a vacancy and spin down with a particle in the 
process of diffusion-limited annihilation the vector $\ket{m,n}$ is the state 
with two particles located at sites $m,n$ and empty sites everywhere 
else \cite{Schu00}. Hence the calculation of the two-point correlation function
is reduced to a two-particle problem of annihilating particles.

The two-time autocorrelation function at site $n=0$ is given by (\ref{2-5})
in terms of the equal-time correlation function. By reflection
symmetry one has $C_{-m,0}(t)=C_{0,m}(t)$ and therefore we may write
\bel{3-5}
C_0(t_w;\tau) =  1 - 2 \sum_{m=1}^\infty 
\mbox{e}^{-\tau}I_m(\tau)[1-C_{0,m}(t_w)].
\ee
With (\ref{3-4}) one has
\bel{3-6}
1-C_{k,l}(t) = \sum_{x=-\infty}^\infty
\sum_{y=x+1}^\infty \bra{s}
\left(1- \prod_{k=-L}^L \sigma^z_k\right)\ket{x,y} P(x,y;t|k,l;0)
\ee
where the two-particle propagator
\bel{3-7}
P(x,y;t|k,l;0) = \mbox{e}^{-2t}
\left[ I_{k-x}(t)I_{l-y}(t)- I_{k-y}(t)I_{l-x}(t)\right]
\ee
for DLPA is the probability that two independent random walkers who started 
at $k,l$ have reached sites $x,y$ after time $t$ without having met on the 
same site. This yields
\bel{3-8}
1-C_{0,m}(t) = 2\sum_{y=-L}^{L} \mbox{e}^{-2t} I_y(t)
\sum_{x=L+1-m}^{L+m}I_x(t).
\ee

We are interested in the derivative $\partial C(t,t_w)/\partial t_w$ 
for large $t$. For an initial size $2L+1$ of the domain of flipped spins
we consider the regime where $t_w \gg L^2$ since at earlier times fluctuations 
in the initial positions of the domain walls have not reached the origin and 
hence would lead to trivial behaviour. To calculate the derivative of the
correlator we express the exact expression (\ref{3-5}) in terms of the
variables $t_w,t$ as in (\ref{2-11bis}). Some algebra similar to the previous 
sections yields for fixed $t$
\bel{3-9}
- \frac{\partial}{\partial t_w} C_0(t,t_w) = 
\frac{2L+1}{\sqrt{2\pi^3 t_w^3}}
\left(\frac{1}{\sqrt{2}\alpha}+\arctan\left(\sqrt{2}\alpha\right)\right).
\ee
With (\ref{2-28}), (\ref{2-30}) the autoresponse function is given by
\bea
R(t_w;\tau) & = & G_0(\tau) \frac{1}{2}(2-C_{0,1}-C_{-1,0}(t_w)) \nonumber \\
\label{3-10}
& = & G_0(\tau) \left[ G_L(t_w) + G_{L+1}(t_w)\right] \sum_{y=-L}^L 
G_{y}(t_w) ~~~~~~
\eea
where we used (\ref{3-6}). In the regime $t_w\gg L^2$
this reduces to 
\bel{3-11}
R(t_w;\tau) = \frac{2L+1}{\sqrt{2\pi^3}} \frac{1}{\sqrt{2}\alpha}
\cdot t_w^{-3/2}.
\ee

As expected both the autocorrelation function and the autoresponse function 
contain the nonuniversal amplitude $2L+1$ which is the initial size of the 
flipped domain. However, the FDR is a universal 
function of the scaling variable $x=1+2\alpha^2$. We find
\bel{3-12}
X(t,t_w) = \hat{X}(x)= \frac{1}{1+\sqrt{x-1}\arctan{(\sqrt{x-1})}}.
\ee
This scaling function is different from (\ref{2-38}), in particular
\bel{3-13}
X_{\infty} = 0
\ee
in contradiction to the universality hypothesis for critical dynamics
as formulated in \cite{Godr00}.

It might be helpful to restate our results in terms of scaling functions. 
In the ageing regime, where both $t_w$ and $t-t_w$ become large, one expects
{\em at} criticality, see e.g. \cite{Godr02}
\bel{3-13a}
C(t,s) = s^{-a} f_C(t/s) \;\; , \;\;
R(t,s) = s^{-1-a} f_R(t/s)
\ee
where $a$ is a non-equilibrium exponent and 
such that for large arguments $x\to\infty$, 
$f_{C,R}(x)\sim x^{-\lambda_{C,R}/z}$ which defines the autocorrelation 
and autoresponse exponents $\lambda_{C}$ and $\lambda_R$, respectively. 
This implies that $\hat{X}(x)\sim x^{(\lambda_C-\lambda_R)/z}$ for $x\to\infty$.
The dynamical exponent $z=2$ throughout in the model at hand but the
other exponents depend on the initial conditions as follows. If we take
an initial state with decaying correlation of the power-law form (\ref{2-19}), 
we read off from the results of section 2
\bel{3-14}
a = 0 \;\; , \;\; \lambda_C = 1 = \lambda_R 
\ee
However, for an initial state of the form (\ref{3-1}), we find
\bel{3-15}
a = \frac{1}{2} \;\; , \;\; \lambda_C = 0 \;\; , \;\;
\lambda_R =1
\ee
We therefore see explicitly that the different forms of the scaling function
$\hat{X}(x)$ signal two distinct dynamical universality classes. 

\section{C-violation and nonuniversality}
\setcounter{equation}{0}

In the previous sections we considered special spin flip dynamics which in 
zero field reduce to Glauber dynamics. The most general flip rates for a local 
magnetic field which (i) satisfy detailed balance with respect to the 
equilibrium distribution (\ref{2-27}), (ii) correspond to reflection-symmetric 
finite-range interactions, and (iii) are nonconserved and in particular 
lead to Glauber rates for $h=0$ are as follows:
\bea
\uparrow\uparrow\uparrow\quad\to\quad\uparrow\downarrow\uparrow & 
\mbox{\rm with rate} & 
(1-\gamma)f_1 \mbox{e}^{-h/T} \nonumber \\
\uparrow\downarrow\uparrow\quad\to\quad\uparrow\uparrow\uparrow &  & 
(1+\gamma)f_1 \mbox{e}^{h/T} \nonumber \\
\uparrow\uparrow\downarrow\quad\to\quad\uparrow\downarrow\downarrow & & 
f_2 \mbox{e}^{-h/T} \nonumber \\
\uparrow\downarrow\downarrow\quad\to\quad\uparrow\uparrow\downarrow & & 
f_2 \mbox{e}^{h/T} \nonumber \\
\downarrow\uparrow\uparrow\quad\to\quad\downarrow\downarrow\uparrow & & 
f_2 \mbox{e}^{-h/T} \nonumber \\
\downarrow\downarrow\uparrow\quad\to\quad\downarrow\uparrow\uparrow & & 
f_2 \mbox{e}^{h/T} \nonumber \\
\downarrow\uparrow\downarrow\quad\to\quad\downarrow\downarrow\downarrow & & 
(1+\gamma)f_3 \mbox{e}^{-h/T} \nonumber \\
\downarrow\downarrow\downarrow\quad\to\quad\downarrow\uparrow\downarrow & & 
(1-\gamma)f_3 \mbox{e}^{h/T} \nonumber 
\eea
Here $f_i=f_i(h/T)$ which may also depend on the neighbouring spin
variables and must be such that $f_i(0)=1$.
With these functions the flip rates at site 0 may be written
\bea
w(h) & = & \left[1-\frac{\gamma}{2}s_0\left(s_{-1}+s_1\right)\right] 
\mbox{e}^{-h/T s_0} \times \nonumber \\
 & & \left[f_1 + 2 f_2 + f_3 + (f_1-f_3)\left(s_{-1}+s_1\right) + 
(f_1 - 2 f_2 + f_3) s_{-1}s_1 + \mbox{\sc srt} \right] \nonumber \\
\label{4-1}
 & = & \left[1-\frac{\gamma}{2}s_0 \left(s_{-1}+s_1\right)\right] 
\mbox{e}^{-h/T s_0} \left[g_0 + g_1 \left(s_{-1}+s_1\right) + 
g_2 s_{-1}s_1 + \mbox{\sc srt} \right].
\nonumber \\
 & & ~
\eea
The short range terms \mbox{\sc srt} 
(which vanish for pure nearest-neighbour interactions) 
involve lattice sites at distances $|k|>1$ from the origin.
Glauber dynamics \cite{Glau63} corresponds to $g_0=1$, $g_1=g_2=0$. 
With these rates the
Markov generator $H$ takes the form $H=\sum_n h_n$ with the local spin flip
generators $h_n$ defined by (\ref{2-7}) for $n\neq 0$ and
\bel{4-2}
h_0 = \frac{1}{2}(1-\sigma^x_0) \hat{w}(h).
\ee
Here $\hat{w}(h)$ is the diagonal matrix obtained from (\ref{4-1}) by 
replacing
the classical spin variables $s_i$ by the Pauli matrices $\sigma^z_i$ 
\cite{Schu00}.

To first order in $h$ the perturbation $V$ of the stochastic time evolution 
that corresponds to this
general choice of rates is obtained from the expansion of $\hat{w}(h)$ to first 
order in $h$. This leads to the following 
general form of the response function
\bel{4-3}
R(t_w;\tau) = G_0(\tau)\left\langle\left[s_0-\frac{\gamma}{2}
\left(s_{-1}+s_1\right)\right]
\left[s_0 - g_0' - g_1' \left(s_{-1}+s_1\right) -g_2's_{-1}s_1 - 
\dots\right]\right\rangle
\ee
where the correlation function is evaluated at time $t_w$.
The dots denote contributions from the \mbox{\sc srt} 
next-nearest neighbour interactions and the constants $g_i'$ are defined by
$g_i'=T\partial g_i/\partial h|_{h=0}$.
Glauber dynamics as used in the previous sections corresponds to the 
choice $g_i'=0$ \cite{Glau63}. The heat bath dynamics of \cite{Godr00} 
corresponds to the choice $g_0'=g_2'=0$, $g_1'=-\gamma/2$.

\subsection{C-invariance}

In the context of spin systems the stochastic dynamics defined by the rate 
functions $f_i$ must remain invariant under simultaneous reversal of all spins 
$s_k \to -s_k$ and change of sign in the external magnetic field. 
This global symmetry is an automorphism on the state space analogous to 
charge conjugation in the quantum field theory of elementary particles.
Because of this analogy, we shall refer to it as  
{\em C-symmetry}. C-symmetry of the rates requires the following properties
\bea
\label{4-4}
f_1(h) & = & f_3(-h) \\
\label{4-5}
f_2(h) & = & f_2(-h) 
\eea
of the rate functions. Therefore $g_0'(0) = g_2'(0) = 0$ for C-invariant
systems.

For translation-invariant initial distributions the response function 
(\ref{4-3}) at 
$T=0$ (corresponding to $\gamma=1$) then takes the form
\bel{4-6}
R^{(C)}(t_w;\tau) = G_0(\tau)\left[1- C_1(t_w)+g_1'\frac{\D}{\D t_w} C_1(t_w)-
\exval{(s_0-\frac{1}{2} \left(s_{-1}+s_1\right))
(\dots)}\right]
\ee
where eqs.~(\ref{2-1},\ref{2-7}) were used. 
The dots denote further \mbox{\sc srt} terms coming from unspecified 
non-nearest-neighbour short range interactions.
For long times the first contribution (proportional to $1-C_1(t)$)
contains no nonuniversal parameter of the dynamics. This is indeed the leading 
contribution since the second contribution, being a time derivative with
respect to $t_w$, is subleading. Finally, the third term is a second-order 
(lattice) partial space derivative which by dynamical 
scaling is of the same subleading order in time as a time-derivative. 

\subsection{Consequences for non-universality} 

The C-symmetry is a physical requirement that any spin dynamics must satisfy. 
However, the Ising model may also be regarded as a classical lattice 
gas model, see e.g. \cite{Kubo65,Chop98}.
In this interpretation the equilibrium distribution (\ref{2-27}) describes 
a system of hard-core particles with attractive nearest-neighbour interactions
which may occupy the sites of a lattice, as indicated by the
occupation numbers $n_i = (1-s_i)/2 \in \{0,1\}$. The external magnetic field
then corresponds to a chemical potential and a spin-flip corresponds to 
particle/vacancy interchange, i.e., an exchange of particles with an external
reservoir. Clearly, in this interpretation of the same model
there is no need to enforce the constraints 
(\ref{4-4}), (\ref{4-5}). Hence it is interesting to investigate universality 
in the absence of this symmetry.

The response function contains the same universal part $R^{(C)}$ from 
eq.~(\ref{4-6}) as in the symmetric case plus two further terms
\bea
\lefteqn{R(t_w;\tau) = R^{(C)}(t_w;\tau)} 
\nonumber \\
& &- G_0(\tau)\left[(1-\gamma)m_0 g_0'- 
\left(\gamma m_0- \exval{s_1(t_w)s_0(t_w)s_1(t_w)}\right)g_2'\right].
\eea
The first term in the second line vanishes at $T=0$ since then $\gamma=1$. 
The second term is non-zero only for symmetric initial states with $m_0=0$. 
As a function of time it is of the same order as $R^{(C)}(t_w;\tau)$. 

\section{Conclusions}

Our study of universality in the critical, nonconserved, 
dynamics of purely classical quenched lattice
models has uncovered some of the basic mechanisms behind the universality of
the scaling of correlation and response functions. 
Explicit calculations for $1D$ kinetic spin systems with non-conserved dynamics 
and which generalize the
Glauber-Ising model have shown that the asymptotic form of the two-time 
response function does {\em not} depend on the microscopic form of the 
stochastic dynamics, provided only that C-symmetry holds.\footnote{For the
$1D$ Glauber-Ising model, we have explicitly shown that our realization of
Glauber dynamics reproduces the same results as the heat-bath dynamics 
studied in \cite{Godr00,Lipp00}.} 
Furthermore, in the scaling limit $t_w\gg 1$ and $t-t_w\gg 1$ 
both the response function and the time-derivative of the 
associated two-time correlation are largely independent of the form of
of the correlations in the initial state, in agreement with the expected
universality of the fluctuation-dissipation ratio $X=\hat{X}(x)$ and its
limit values $X_{\infty}=\lim_{x\to\infty}\hat{X}(x)$. This universality holds 
true with respect to the long-range decay of the initial 
correlations (see eq.~(\ref{1-3}) with $\nu\geq 0$) 
and also with respect to the `details' of the dynamics. On the other hand, 
we have also shown the existence of two {\em distinct} kinetic
universality classes which arise for initial magnetization $|m_0|=1$ and 
$|m_0|<1$, respectively.

In the absence of C-invariance these results only hold true for 
symmetric initial states with $m_0=0$. For $m_0\ne 0$, a non-universal
part will contribute to the long-time behaviour of two-time correlation
and response functions. Since lattice gases are in general not C-invariant,
a study of these system will make such terms apparent. In this context,
recent attempts \cite{Chat03,Ricc03} to construct algorithms which allow
to measure the two-time response directly, may be of value. 

Finally, the available exactly solvable models (the $1D$ Glauber-Ising model
and the kinetic mean spherical model) have revealed two distinct possible
routes towards modifications of critical dynamics beyond a fully disordered
initial state: (1) the presence of large ordered domains and (2) the interplay
of strong initial fluctuations with the thermal fluctuations of the bulk. 
However, in these models, the two-points functions decouple and can be
studied independently of any longer-ranged correlations. A test of our scenario
of possible non-universalities in more general systems, such as the
$2D$ Glauber-Ising model, is called for.

\zeile{2}

\noindent 
{\bf Acknowledgements:} MH thanks the Centro de F\'{\i}sica Te\'orica e 
Computacional (CFTC) of the Universidade de Lisboa for warm hospitality. 
GMS thanks the LPM in Nancy for kind hospitality and providing a 
stimulating environment.


\end{document}